\documentclass{llncs}
\usepackage{makeidx}  
\usepackage{graphicx}
\usepackage{amsmath}
\usepackage{multirow}
\usepackage{algorithm}
\usepackage{algorithmic}
\usepackage{url}

\begin{document}
\mainmatter              
\title{Synthetic sequence generator for recommender systems - memory biased random walk on sequence multilayer network}
\titlerunning{Synthetic clickstream generation algorithm}  
%
\author{Nino Antulov-Fantulin\inst{1} \and Matko Bo\v{s}njak\inst{1} 
 \and Vinko Zlati\'{c}\inst{2} \and Miha Gr\v{c}ar\inst{3} \and Tomislav \v{S}muc\inst{1} }
\authorrunning{Nino Antulov-Fantulin et al.} 
%
\tocauthor{Nino Antulov-Fantulin, Matko Bo\v{s}njak, Vinko Zlati\'{c} and Tomislav \v{S}muc}
\institute{Laboratory for Information Systems, Division of Electronics, Rudjer Bo\v{s}kovi\'{c}  Institute, Zagreb, Croatia,\\
\email{nino.antulov@irb.hr},\\ 
\and
Theoretical Physics Division, Rudjer Bo\v{s}kovi\'{c} Institute, Zagreb, Croatia
\and 
Department of Knowledge Technologies / E8, Jo\v{z}ef Stefan Institute, Ljubljana, Slovenia
}

\maketitle              

\begin{abstract}
Personalized recommender systems rely on each user's personal usage data in the system, in order to assist in decision making.
However, privacy policies protecting users' rights prevent these highly personal data from being publicly available to a wider researcher audience.
In this work, we propose a memory biased random walk model on multilayer sequence network, as a generator of synthetic sequential data for recommender systems.
We demonstrate the applicability of the synthetic data in training recommender system models for cases when privacy policies restrict clickstream publishing.
\end{abstract}

\section{Introduction}
Recommender systems provide a useful personal decision support in search through vast amounts of information on the subject of interest \cite{StateArtRS1}~\cite{StateArtRS}, such as books, movies, research papers, etc.
The operation and performance of recommender systems based on collaborative data \cite{hybrid1}~\cite{colFilt2}
 are necessarily tied to personal usage data, such as users' browsing and shopping history, and to other personal descriptive data such as demographical data.
These data often conform to privacy protection policies, which usually prohibit their public usage and sharing, due to their personal nature.
This, in turn, limits research and development of recommender systems to companies in possession of such vital data, and thus prevents performance comparison of new systems between different research groups.
In order to enable data sharing and usage, many real data sets were anonymized by removing all the explicit personal identification attributes like names and social security numbers, among others.
Nevertheless, various research groups managed to successfully identify personal records by linking different datasets over quasi-personal identifiers such as search logs, movie ratings, and other non-unique data, revealing as a composition of identifiers.
Due to successful privacy attacks, some of the most informative data for recommendation purposes, such as the personal browsing and shopping histories, are put out of the reach of general public.
In their original form, usage histories are considered personal information, and their availability is heavily restricted.
However, even with the personal information obfuscated, they remain a specific ordered sequence of page visits or orders, and as such can be uniquely tied to a single person through linkage attacks.
With usage histories often rendered unavailable for public research, recommender systems researchers have to manage on their own and often work on disparate datasets.
Recently, a one million dollar worth Overstock.com recommender challenge released synthetic data, which shares certain statistical properties with the original dataset. The organizers noted that this dataset should have been used only for testing purposes, while the code itself had to be uploaded to RecLabs\footnote{\url{http://code.richrelevance.com/
reclab-core/}} for model building and evaluation against the real data. The challenge ended with no winner since no entry met the required effectiveness at generating lift. It would be useful both for contestants and the companies, if the synthetic data could be used for recommendation on real users.

Random walks \cite{RS_RW}, \cite{MovieRS_RW}, \cite{WebRS_RW} have been used for constructing recommender systems on different types of graph structures originating from users' private data, but not to generate synthetic clickstreams. We propose an approach to synthetic clickstream generation by constructing a memory biased random walk model (MBRW) on the graph of the clickstream sequences, which is a subclass of Markov chains \cite{ProbFeller}, \cite{ProbBook1}. We show that the synthetic clickstreams share similar statistical properties to real clickstream. We also use the MBRW model to generate synthetic clickstreams for the VideoLectures.NET\footnote{\url{http://videolectures.net}} dataset from the ECML/PKDD 2011 Discovery Challenge\cite{antulov2011ecml}. And finally, we demonstrate that synthetic data could be used to make recommendations to real users on the Yahoo Music dataset released for the KDD\-Cup challenge for the year 2011 \cite{KDDCup2011}.

\section{Methodology}

The biased random walk on a graph \cite{RW_vinko}~\cite{NetworksBook} is a stochastic process for modelling random paths on a general graph structure. Clickstream is a sequence of items (path on graph), such as web pages, movies, books, etc., a user interacted with, i.e.~user's interaction history. Throughout the rest of this work we assume, without the loss of generality, a clickstream as a sequence of visited web pages. Formally, a clickstream $c^i$ is defined as an ordered sequence of web pages $c^i=\{u_1^i, u_2^i, u_3^i, ... , u_n^i\}$, visited by a particular user $i$. The set of all the clickstreams in a system is $C=\{c^1,c^2, ... ,c^i,..., c^m\}$.
Now, we describe the memory biased random walk which will be used to generate sequences of synthetic paths on specific networks.  
The two characteristic data generator matrices used in this work are the Direct Sequence matrix ($DS$) and the Common View Score matrix ($CVS$).
The element $DS[m,n]$ of the matrix $DS$ denotes the number of clickstreams in $C$ in which the web page $m$ immediately follows the web page $n$.
The element $CVS[m,n]$ of the matrix $CVS$ denotes the number of occurrences in which the web page $m$ and the web page $n$ belong to the same clickstream in $C$.
Using these generator matrices, we can now define the memory biased random walk on multilevel network. 
By introducing the memory component to the biased random walk \cite{RW_vinko} , we obtain the MBRW model. The MBRW model is a discrete time Markov chain model, with a finite memory of $m$ past states. Biases from the $DS$ graph are connecting probability of choosing next item and current item, while, biases from the $CVS$ graph are connecting probability of choosing the next item with the past $m$ items in a clickstream.
The initial vertex for the random walk can be chosen by either a stochastic or a deterministic rule.
Given an initial vertex $u_1$, the probability of choosing the adjacent vertex $u_2$ equals:
\begin{equation}
P_{\{u_2|u_1\}} = \frac{ DS_{u_2,u_1}} {\sum_{k} DS_{k,u_1} }
\end{equation}
\noindent which, in turn, generates a clickstream $c^i=\{u_1,u_2\}$.
The third vertex, $u_3$ in the clickstream is chosen with the probability of:
\begin{equation}
P_{\{u_3|u_2,u_1\}} = \frac{ DS_{u_3,u_2} CVS_{u_3,u_1} } {\sum_{k} DS_{k,u_1} CVS_{k,u_1}} 
\end{equation}
\noindent thus generating a clickstream $c^i=\{u_1,u_2,u_3\}$.
Using a finite memory of size $m$, we choose the vertex $u_n$ with the probability of:
\begin{equation}
\label{mbrwFormula}
P_{\{u_n|u_{n-1},...,u_{n-m-1}\}} = \frac{ DS_{u_n,u_{n-1}}  \prod_{k=1}^{m} CVS_{u_n,u_{n-k-1}} } 
{\sum_{j} DS_{j,u_{n-1}} \prod_{k=1}^{m} CVS_{j,u_{n-k-1}} } 
\end{equation}
\noindent thus generating a clickstream $c^i = \{u_1,u_2,u_3,...,u_n\}$ at the $n$-th step of the random walk.

\begin{figure}[htpb]
	\begin{center}
		\setlength\fboxsep{0.5pt}
		\setlength\fboxrule{0.5pt}
		\includegraphics[width=80mm]{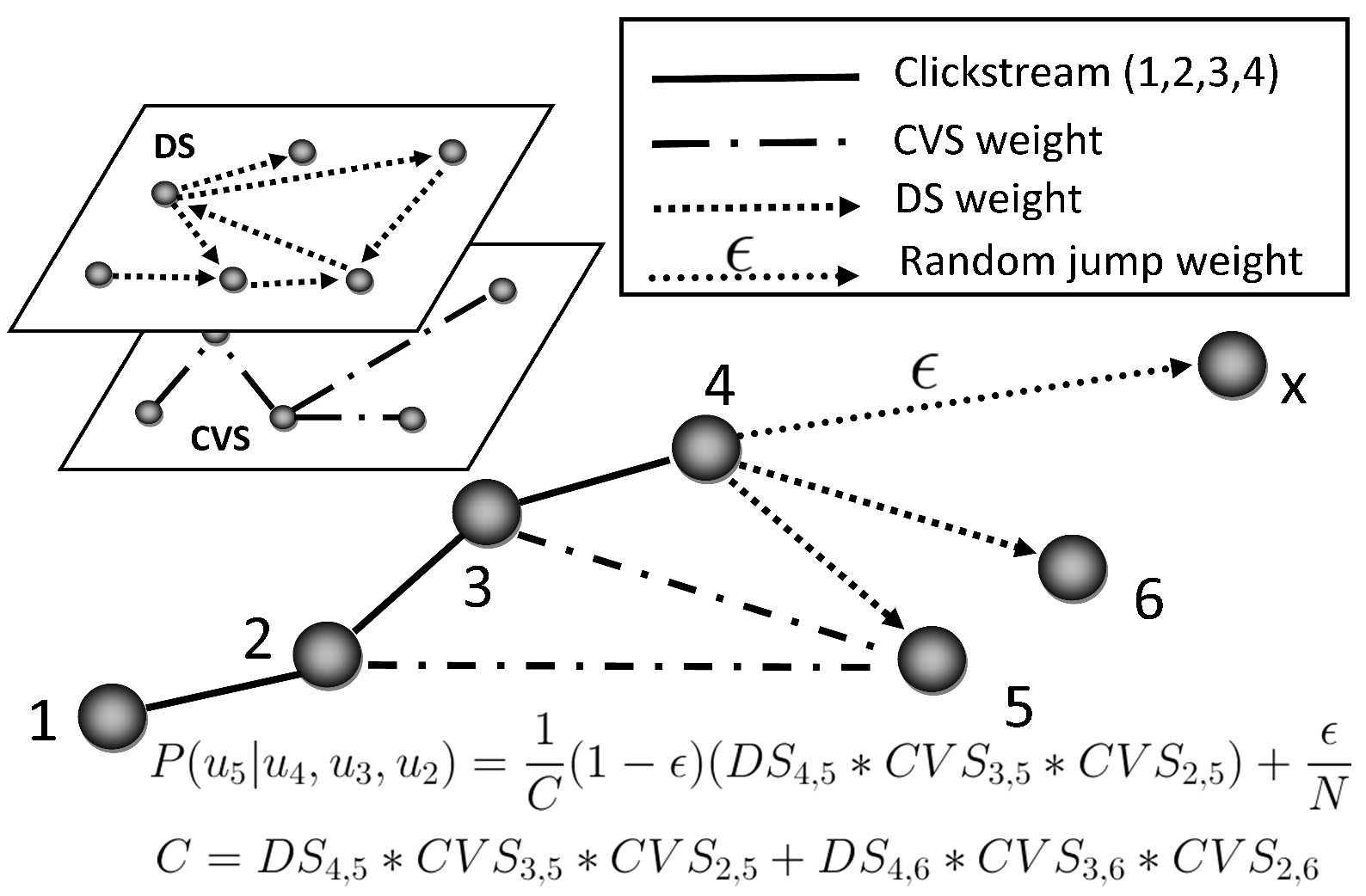}
		\caption{ Simple example: at current step the MBRW model ($m=2$) has created a clicktream $(u_1,u_2,u_3,u_4)$ and node $u_4$ has two neighbouring nodes $u_5$ and $u_6$ at the $DS$ graph. The transition probability (see formula \ref{mbrwFormula}) to node $u_5$ is given, where the $\epsilon$ transition denotes the probability of a jump to some arbitrary node $u_x$, the $C$ denotes the normalization, the factor $\frac{\epsilon}{N}$ denotes the probability of random jump back to node $u_5$ and $N$ denotes total number of nodes in $DS$ graph.}
		\label{fig:MBRW_formula_vis}
	\end{center}
\end{figure}

The intuition behind this analytical expression for transition probability is the following. The probability of choosing the next item is proportional to the product of direct sequence frequency $DS$ and common view score frequencies $CVS$ in clickstream data. Direct sequence frequency $DS$ measures the tendency that the current item appears one place before the next item in clickstream data. Product of common view score frequencies tell us the tendency that the next item appears with all other items in current clickstream that we are generating. Denominator is the normalization expression to turn frequencies to probability measure. In Figure \ref{fig:MBRW_formula_vis}, we demonstrate how transition probability is calculated on a simple example.
Clickstream length $L$ is a random variable sampled from a discrete probability distribution like Poisson, negative binomial, geometric, or from the real clickstream length distribution, if available. Using this model, we generate a set of synthetic clickstreams $C^{*}=\{c^{*1},c^{*2},...,c^{*K}\}$. In each of the $K$ independent iterations, we determine the clickstream length $l$ and the initial vertex of the random walk. At the end of each iteration $i$, random walk path $c^i=\{u_1^i,u_2^i,...,u_l^i\}$ determines one clickstream appended to the synthetic clickstream set $C^{*}$.

In order to ensure an additive smoothing over transition probabilities in MBRW walk, we introduce a small $\epsilon$ probability of a random jump. At each step in the clickstream generation process a random walker produces a jump to some random item with the probability $\epsilon$. This $\epsilon$-smoothing technique turns all possible clickstreams to become non-forbidden in generation process. The pseudo code for Memory Biased Random Walks with Random Jumps is provided in Algorithm \ref{alg2}. The code for the MBRW model is available at github repository\footnote{\url{http://github.com/ninoaf/MBRW}}.

\begin{algorithm}
\caption{Memory Biased Random Walks with Random Jumps}
\label{alg2}
\begin{algorithmic}
\STATE \textbf{Input:} 
$DS$ - Direct Sequence matrix, $CVS$ - Common View Score matrix,
$K$ - number of synthetic clickstreams, 
$\epsilon$ - probability of random jump, 
$m$ memory length from prob. distr. $M$
\STATE \textbf{Output:} $ C^{*}=\{ c^{*1},c^{*2},...,c^{*K}\}$ synthetic clickstream set
\STATE $ C^{*}= \emptyset$ 
\FOR{$i$ = 1 : K }
\STATE $c^{*}_i = u$ // random initial item;
\FOR{$j$ = 1 : number of random hops }
\STATE with $1-\epsilon$ probability choose the next item $u_{j}$ with MBRW walk on $DS$ and $CVS$;
\STATE with $\epsilon$ probability choose the next item $u_{j}$ with random jump;
\STATE append new item: $c^{*}_i = c^{*}_i \cup u_{j}$;  
\ENDFOR
\STATE append new synthetic clickstream: $ C^{*}= C^{*} \cup c_i$;
\ENDFOR
\end{algorithmic}
\end{algorithm}


\section{Results}
We analyse the statistical properties as well as the utility of the synthetic data in training recommender system models. In our experiments we will use two datasets:
(i) Yahoo Music dataset released for the KDD\-Cup challenge for the year 2011 \cite{KDDCup2011} and (ii) the VideoLectures.NET\footnote{\url{http://videolectures.net}} dataset from the ECML/PKDD 2011 Discovery Challenge\cite{antulov2011ecml}. 
As the privacy policies did not restrict publishing user preferences data to particular musical items in KDD\-Cup challenge 2011, the subsample of the Yahoo Music dataset is used in our study as a experimental polygon to measure the ability of training the recommender systems models on synthetic data. Note, that all the content data about musical items were anonymized in the Yahoo Music dataset. Contrary, in the ECML/PKDD 2011 Discovery Challenge \cite{antulov2011ecml}, only the content data and clickstrem statistics could be published but not the actual clickstreams. Therefore, we use our methodology on the second VideoLectures.NET dataset to create and publish synthetic clickstream data.
The first dataset used in our experiments is a subset of Yahoo Music dataset released for the KDD\-Cup challenge for the year 2011 \cite{KDDCup2011}
Yahoo 2011 challenge dataset which contains user preferences (ratings) to particular musical items along with appropriate time stamp. We extracted from this dataset a subset that represents a very good proxy for a set of sequential activity (clickstreams). For each user in our subset we retained sequence of highly rated items in ascending order over time stamps (sequence activity or clickstream proxy). We limited the total number of items and users in our subset to 5000 and 10000 respectively, in order to be able to perform large set of computational experiments with resources on disposal. The reduced dataset is denoted with $C$ represents a set of clickstreams for 10000 users. This dataset reduction should not have any significant impact on the results and conclusions of the study. We will address this question later with cross-validation technique. 

Our first hypothesis is the following. Given a sufficiently large synthetic dataset, basic statistical properties of $DS^{*}$ and $CVS^{*}$ matrices are preserved. We examined how statistical properties of the item preference matrix like $DS$ and $CVS$ are preserved in synthetic clickstream set, with respect to the original clickstream set. We calculated the $DS$ and $CVS$ matrices from the $C$ dataset and created the synthetic clickstream set $C^{*}$ by using the MBRW model. Memory parameter $m$  was sampled from the Gaussian distribution $\mathcal{N}(3,2^2)$, number of random walk hops parameter $l$ was sampled from $\mathcal{N}(9,2^2)$ and number of synthetic clickstreams parameter $K$ varying from $10^4-10^6$.
Upon obtaining the synthetic clickstream set $C^{*}$, we calculated the $DS^{*}$ and $CVS^{*}$ matrices, and compared their statistical properties to the original matrices $DS$ and $CVS$. We have used the Spearman's rank correlation measure between the corresponding rows in ($DS,DS^{*}$) and ($CVS,CVS^{*}$). 

\begin{table}[H]
\caption{Average rank correlation between ($DS$,$DS^{*})$ and ($CVS$,$CVS^{*}$) for different sizes (K) of generated synthetic clickstream set. Synthetic clickstream set is created using parameter m sampled from $\mathcal{N}(3,2^2)$, parameter l sampled from $\mathcal{N}(9,2^2)$. 
}
\begin{center}
    \begin{tabular}{ | l | l | l | } \hline
 	Size & $AVG[r(DS,DS^{*})]$  & $STD[r(DS,DS^{*})]$ \\ \hline
    $K=10^4$ & 0.5700 & 0.3210 \\ \hline
    $K=5*10^4$ & 0.8261 & 0.3060\\ \hline
    $K=10^5$ & 0.8914 & 0.2224 \\ \hline
    $K=5*10^5$ & 0.9308 & 0.0639 \\ \hline
    $K=10^6$ & 0.9294 & 0.0590\\ \hline
     & $AVG[r(CVS,CVS^{*})]$  & $STD[r(CVS,CVS^{*})]$ \\ \hline
    $K=10^4$ & 0.4545 & 0.2677 \\ \hline
    $K=5*10^4$ & 0.5530 & 0.2407\\ \hline
    $K=10^5$ & 0.6050 & 0.2120 \\ \hline
    $K=5*10^5$ & 0.7071 & 0.1765 \\ \hline
    $K=10^6$ & 0.7361 & 0.1784\\ \hline
    \end{tabular}
\end{center}
\label{tab:tab1} 
\end{table}

Due to the fact that these matrices are sparse and that in the process of recommendation only top ranked items are relevant, we limited the rank correlation calculation to the first $z=100$ elements. Rank correlation between complete rows would be misleadingly high due to the row sparsity.
Average rank correlation coefficient $AVG[r(DS,DS^{*})]=0.92$ and $AVG[r(CVS,CVS^{*})]=0.73$ over all corresponding rows was obtained for the first $z$ most important elements, with the above parameters and $K=10^6$. The rank correlation coefficients for different values of parameter $K$ can be seen in Table \ref{tab:tab1}.
This shows highly correlated statistical properties ($DS,DS^{*}$) and ($CVS,CVS^{*}$).

Now, we analyse the ability to learn recommender system models from synthetic data and apply this model on real users. We measure and compare the recommender system models learned on real, synthetic and random data and their corresponding performance on recommending items to real users. We take the standard Item-Knn \cite{ItemKnn} recommender system as a representative of similarity-based techniques and state-of-the-art method matrix factorization techniques: Bayesian Personalized Ranking Matrix Factorization Technique \cite{BPRMF}. We hypothesise that learning recommender systems models even from synthetic data can help making predictions to real users. 

\begin{figure}
	\begin{center}
		\setlength\fboxsep{0.5pt}
		\setlength\fboxrule{0.5pt}
		\includegraphics[width=100mm]{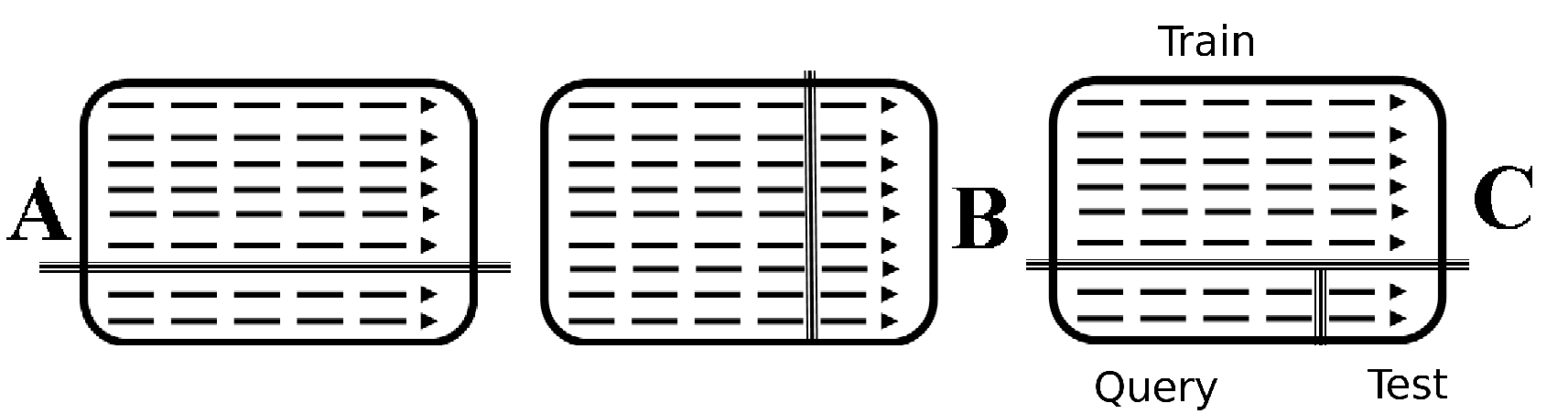}
		\caption{Three ways of splitting the original clickstream set used in computational experiments: A - Horizontal split, B - Vertical split and C - Horizontal and vertical split}
		\label{fig:splits}
	\end{center}
\end{figure}

\begin{figure}
	\begin{center}
		\setlength\fboxsep{0.5pt}
		\setlength\fboxrule{0.5pt}
		\includegraphics[width=120mm]{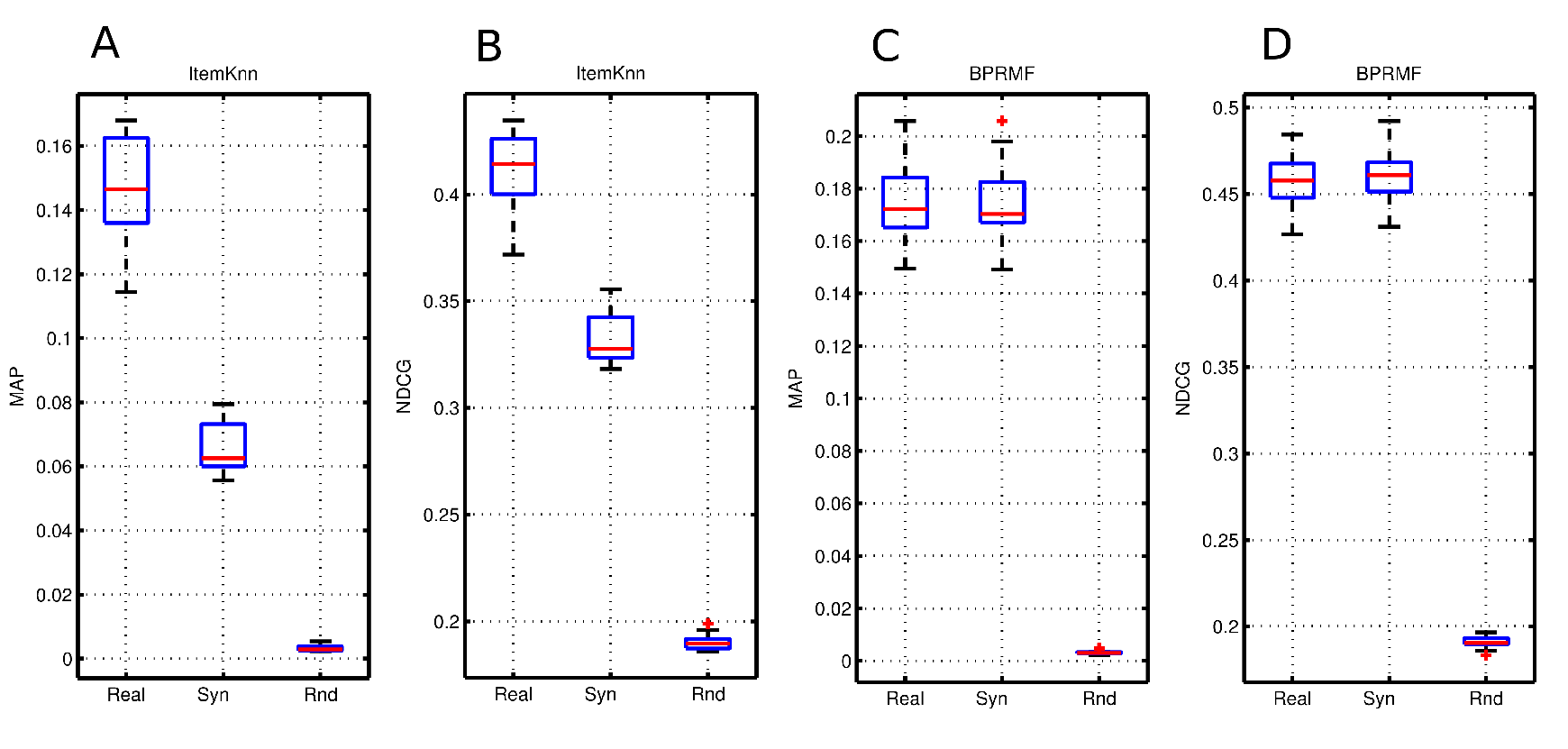}
		\caption{Results for 10-folds cross-validation for MAP and NDCG measures for different datasets with Item-Knn \cite{ItemKnn} (plots A and B) and BPRMF \cite{BPRMF} algorithm (plots C and D). Label "Real" represents performance on real dataset. Label: "Syn" represents synthetic data using MBRW with $m$ sampled from  $\mathcal{N}(3,2^2)$, $l$ sampled from real length distribution, $\epsilon=0.0001$, $K$=9000. Label "Rnd" represents random data generated by random jumps $\epsilon=1.0$ on item graph. }
		\label{fig:CrossValid}
	\end{center}
\end{figure}

In order to create proper training and test data for testing of our hypotheses we create two splits: a "vertical" and "horizontal" split. The horizontal split of the clickstream dataset $C$ randomly divides them to two disjoint, fixed-size clickstream sets $C_{train}$ and $C_{test}$. 
Using the horizontal split on our Yahoo! dataset, we produced a training set $C_{train}$ containing 9000, and a test set $C_{test}$ containing 1000 clickstreams. 
The vertical split, on the other hand, divides clickstream in $C$ into two sets: first $t$ items are appended to first set, whereas the rest of the clickstream items belong to a second set. By applying the vertical split on the $C_{test}$ set again, we get two additional sets $C_{query}$ (first $50\%$ of items in a clickstream) and $C_{test}$ (rest of clickstreams). These splits are graphically represented in the Figure \ref{fig:splits}. Experimental procedure is the following.
We extract $DS$ and $CVS$ statistics from $C_{train}$ and generate synthetic $C^*_{train}$ with the MBRW model. The baseline random synthetic dataset $C^*_{RND}$ is created by setting the parameter $\epsilon = 1$ (random jump model). Now, we create three different recommender system models: $M$ (real model), $M^*$ (synthetic model), and $M_{RND}$ (random model) from the $C_{train}$, $C^*_{train}$ and $C^*_{RND}$, respectively. Then recommender models for the input of real users $C_{query}$ produce recommendations which are compared to $C_{test}$ (ground truth). The performance on $C_{test}$ is measured with the standard information retrieval measures: MAP (Mean Average Precision), NDCG (Normalized Discounted Cumulative Gain: ranking measure) and precision@10 (fraction of the top 10 items retrieved by the system that are relevant for the user). In order to estimate how performance results can generalize to independent datasets we use a cross-validation technique. We make a 10-fold horizontal splits of our dataset. Then in each round we generate $C_{train}^i$, $C_{test}^i$ and $C_{query}^i$. For each $C_{train}^i$ we generate synthetic $C^{i*}_{train}$ and random dataset $C^{i*}_{RND}$. Note, that in each round recommender algorithms learn model on $C_{train}^i$, $C^{i*}_{train}$ and $C^{i*}_{RND}$ but their performance is measured for new users $C_{query}^i$ on $C_{test}^i$. In Figure \ref{fig:CrossValid}, we observe that BPRMF and Item-Knn models have significantly better performance than baseline random models. We have used the recommender systems\footnote{Item-Knn with $k$ = 15 and BPRMF with num factors: 10, user, item and negItem regularization: 0.025, iterations: 30, learn rate: 0.05, initial mean: 0.0, initial std: 0.1 and fast sampling: 1024.} implementations from the Recommender System extension \cite{MyMediaLight} \cite{RMplugin} in the RapidMiner.

In the end, we focus on the ECML/PKDD 2011 Discovery Challenge \cite{antulov2011ecml}, where the privacy policies have restricted public availability of users clickstream data on the VideoLectures.Net. This challenge provided a rich content data about items in a system and different statistics about users clickstream sequences. This has motivated us to use the direct sequence statistics and common view statistics as a generators of synthetic clickstreams with a proposed MBRW model. Direct sequence graph $DS$ from this dataset consists of 7226 vertices in a single large, weakly connected component and common view score undirected graph $CVS$ from this dataset consists of 7678 vertices in a large connected component. 
We have produced and published \footnote{\url{http://lis.irb.hr/challenge/index.php/dataset/}} the 20000 synthetic clickstreams for VideoLectures.net with the MBRW model with the memory parameter $m=5$ and clickstream length L sampled from as a Geometric distribution with parameter 0.1 (expected length of clickstreams is 10). 

\section*{Discussion and related work}
The problems of privacy-preserving data publishing \cite{PPDP2}~\cite{PPDP1} and privacy preserving data mining \cite{PPDM} are intensively researched within the database, the statistical disclosure, and the cryptography communities. Recently, a comprehensive survey \cite{Berendt2012} on the privacy challenges and solutions in privacy-preserving data mining has been published.
Different privacy protection models already exists and here we will only mention the important ones.
Record linkage models like k-Anonymity model \cite{Kenig2012KanonyAlg}~\cite{kAnonymity1}~\cite{kAnonymity2} assure that the number of records with a quasi-identifier id is at least $k$ and therefore assure the value of linkage probability at most $1/k$.
Attribute linkage models like L-diversity \cite{lDiversity2} are envisioned to overcome the problem of inferring the sensitive values from $k$ anonymity groups by decreasing the correlations between the quasi-identifiers and the sensitive values.
Probabilistic models like $\epsilon$-differential privacy model \cite{differentialModel} ensures that individual's presence or absence in the database does not effect the query output significantly. Post-random perturbation (PRAM) methods \cite{PRAM2}~\cite{PRAM1} change original values through probabilistic mechanisms and thus, by introducing uncertainty into data, reduce the risk of re-identification.
Aggarwal et.al.~\cite{AggarwalStringAnonymization} proposed an anonymization framework for string-like data.
They used the condensation-based techniques to construct condensed groups and their aggregate statistics.
From the aggregate statistics, they calculated the first and second order information statistics of symbol distributions in strings, and generated synthetic, pseudo-string data. But still, many data-privacy researchers agree that high dimensional data poorly resist to de-anonymization \cite{NetflixAttack} which poses privacy issues for companies, and prevent the usage of real-life datasets for research purposes.
Contrary to standard anonymization methods, synthetic data generation is an alternative approach to data protection in which the model generates synthetic dataset, while preserving the statistical properties of the original dataset. 
Several approaches for synthetic data generation have been proposed: (i) synthetic data generation by multiple imputation method \cite{Synthetic1}, (ii) synthetic data by bootstrap method \cite{Synthetic2} (estimating multi-variate cumulative probability distribution, deriving similar c.d.f., and sampling a synthetic dataset), (iii) synthetic data by Latin Hypercube Sampling \cite{Synthetic3}, (iv) and others such as a combination of partially synthetic attributes and real non-confidential attributes \cite{Synthetic5}~\cite{Synthetic4}. These synthetic data generation strategies were mostly developed for database records with a fixed number of attributes but not for sequences such as sequence data. 

We propose a novel approach for synthetic sequence generation by constructing the memory biased random walk (MBRW) model on the multilevel network of usage sequences.
Moreover, we demonstrate this synthetic data can be used for learning recommender models which can be useful for applications on real users. What are the potential privacy breach problems of our approach ? 
Our method is based on the assumption that the sequence statistics: direct sequence $DS$ and common view score $CVS$ can be publicly available without breaking privacy of particular user. Why this is the case ? We can view the clickstreams as a different way of writing the sequence statistics like finite state machines represent finite way of coding the infinite set of word from some regular language. 
Note, that the privacy breach can occur in a situation when the attacker can claim that individual unique synthetic subsequences could only be generated by using the unique transitions from particular user $u$. This is the reason why we need smoothing procedure ($\epsilon$ jumps) or k-anonymity filtering over the transition matrices $DS$ and $CVS$. The $\epsilon$ random jumps in the generation process with small $\epsilon$ probability correspond to the additive smoothing of transition probabilities in MBRW model. Let us define the set of all possible combinatoric combinations of clickstreams with arbitrary length from set of items with $\Omega$ (infinite). Note, that when $\epsilon = 0$ the MBRW model cannot create arbitrary clickstream from space of all clickstreams combinations $\Omega$ due to the existence of zero values in $DS$ and $CVS$ matrices. As the additive smoothing technique turns all combinatoric clickstreams from $\Omega$ set possible the attacker can not longer certainly claim that certain unique user subsequence was used in generation process. 
K-anonymity filtering can also be applied to $CVS$ and $DS$ directly by filtering all frequencies that are lower than $k$. This filtering enables that the presence or absence of individual transitions in $DS$ or $CVS$ can not be detected. Therefore if the $DS$ and $CVS$ statistics after a potential filtering can be publicly available without breaking privacy our methodology can be applied.


\section{Conclusion}
The principle aim of our work was to construct a generator of real-like clickstream datasets, able to preserve the original user-item preference structure, while at the same time addressing privacy protection requirements.
With respect to this aim, we investigated properties of the memory biased random walk model. 
We demonstrated that the basic statistical properties of data generators $DS$ and $CVS$ matrices are preserved in synthetic dataset if we generate dataset of sufficiently large size. In addition to presenting the MBRW model for synthetic clickstream generation, we demonstrate that the synthetic datasets created with it can be used to learn recommender system models which can be useful to recommendations for real users.


\section{Acknowledgments} 
This work was supported by the European Community $7^{th}$ framework ICT-2007.4 (No 231519) "e-LICO: An e-Laboratory for Interdisciplinary Collaborative Research in Data Mining and Data-Intensive Science" and by the EU-FET project MULTIPLEX (Foundational Research on MULTIlevel comPLEX networks and systems, grant no. 317532).

%
%


\begin{thebibliography}{10}
\providecommand{\url}[1]{{#1}}
\providecommand{\urlprefix}{URL }
\expandafter\ifx\csname urlstyle\endcsname\relax
  \providecommand{\doi}[1]{DOI~\discretionary{}{}{}#1}\else
  \providecommand{\doi}{DOI~\discretionary{}{}{}\begingroup
  \urlstyle{rm}\Url}\fi

\bibitem{StateArtRS1}
Adomavicius, G., Tuzhilin, A.: Toward the next generation of recommender
  systems: A survey of the state-of-the-art and possible extensions.
\newblock IEEE Transactions on Knowledge and Data Engineering \textbf{17}(6),
  734--749 (2005)

\bibitem{AggarwalStringAnonymization}
Aggarwal, C.C., Yu, P.S.: A framework for condensation-based anonymization of
  string data.
\newblock Data Min. Knowl. Discov. \textbf{16}(3), 251--275 (2008)

\bibitem{PPDM}
Aggarwal, C.C., Yu, P.S.: Privacy-Preserving Data Mining: Models and
  Algorithms, 1 edn.
\newblock Springer Publishing Company, Incorporated (2008)

\bibitem{antulov2011ecml}
Antulov-Fantulin, N., Bo{\v{s}}njak, M., Znidar{\v{s}}ic, M., Gr\v{c}ar, M.,
  Morzy, M., \v{S}muc, T.: Ecml/pkdd 2011 discovery challenge overview.
\newblock In: Proceedings of the ECML-PKDD 2011 Workshop on Discovery
  Challenge, pp. 7--20 (2011)

\bibitem{Berendt2012}
Berendt, B.: More than modelling and hiding: towards a comprehensive view of
  web mining and privacy.
\newblock Data Min. Knowl. Discov. \textbf{24}(3), 697--737 (2012)

\bibitem{MovieRS_RW}
Bogers, T.: Movie recommendation using random walks over the contextual graph.
\newblock In: {Second Workshop on Context-Aware Recommender Systems} (2010)

\bibitem{hybrid1}
Burke, R.: Hybrid recommender systems: Survey and experiments.
\newblock User Modeling and User-Adapted Interaction \textbf{12}(4), 331--370
  (2002)

\bibitem{RW_vinko}
Zlati\ifmmode~\acute{c}\else \'{c}\fi{}, V., Gabrielli, A., Caldarelli, G.:
  Topologically biased random walk and community finding in networks.
\newblock Phys. Rev. E \textbf{82}, 066,109 (2010)

\bibitem{PPDP2}
Chen, B.C., Kifer, D., LeFevre, K., Machanavajjhala, A.: Privacy-preserving
  data publishing.
\newblock Found. Trends databases \textbf{2}(18211;2), 1--167 (2009)

\bibitem{Synthetic3}
Dandekar, R.A., Cohen, M., Kirkendall, N.: Sensitive micro data protection
  using latin hypercube sampling technique.
\newblock In: Inference Control in Statistical Databases, From Theory to
  Practice, pp. 117--125 (2002)

\bibitem{Synthetic5}
Dandekar, R.A., Domingo-Ferrer, J., Seb{\'e}, F.: Lhs-based hybrid microdata vs
  rank swapping and microaggregation for numeric microdata protection.
\newblock In: Inference Control in Statistical Databases, From Theory to
  Practice, pp. 153--162 (2002)

\bibitem{ItemKnn}
Deshpande, M., Karypis, G.: Item-based top-n recommendation algorithms.
\newblock ACM Transactions on Information Systems \textbf{22}(1), 143--177
  (2004)

\bibitem{KDDCup2011}
Dror, G., Koenigstein, N., Koren, Y., Weimer, M.: The yahoo! music dataset and
  kdd-cup'11.
\newblock In: Proceedings of KDDCup 2011 (2011)

\bibitem{differentialModel}
Dwork, C.: Differential privacy.
\newblock In: M.~Bugliesi, B.~Preneel, V.~Sassone, I.~Wegener (eds.) Automata,
  Languages and Programming, \emph{Lecture Notes in Computer Science}, vol.
  4052, pp. 1--12 (2006)

\bibitem{ProbFeller}
Feller, W.: An introduction to probability theory and its applications, vol.~2.
\newblock John Wiley \& Sons (2008)

\bibitem{Synthetic2}
Fienberg, S.: A radical proposal for the provision of micro-data samples and
  the preservation of confidentiality.
\newblock Tech. rep., Department of Statistics, Carnegie-Mellon University.
  Technical Report (1994)

\bibitem{WebRS_RW}
Fouss, F., Faulkner, S., Kolp, M., Pirotte, A., Saerens, M.: Web recommendation
  system based on a markov-chain model.
\newblock In: International Conference on Enterprise Information Systems (ICEIS
  2005) (2005)

\bibitem{PPDP1}
Fung, B.C., Wang, K., Fu, A.W.C., Yu, P.S.: Introduction to Privacy-Preserving
  Data Publishing: Concepts and Techniques, 1st edn.
\newblock Chapman \& Hall/CRC (2010)

\bibitem{MyMediaLight}
Gantner, Z., Rendle, S., Freudenthaler, C., Schmidt-Thieme, L.: Mymedialite: A
  free recommender system library.
\newblock In: Proceedings of the fifth ACM conference on Recommender systems,
  pp. 305--308. ACM (2011)

\bibitem{RS_RW}
Gori, M., Pucci, A.: Research paper recommender systems: A random-walk based
  approach.
\newblock In: Web Intelligence, pp. 778--781 (2006)

\bibitem{ProbBook1}
Kao, E.: An introduction to stochastic processes.
\newblock Business Statistics Series. Duxbury Press (1997)

\bibitem{Kenig2012KanonyAlg}
Kenig, B., Tassa, T.: A practical approximation algorithm for optimal
  k-anonymity.
\newblock Data Min. Knowl. Discov. \textbf{25}(1), 134--168 (2012)

\bibitem{lDiversity2}
Machanavajjhala, A., Kifer, D., Gehrke, J., Venkitasubramaniam, M.:
  L-diversity: Privacy beyond k-anonymity.
\newblock ACM Transactions on Knowledge Discovery from Data \textbf{1}(1)
  (2007)

\bibitem{RMplugin}
Mihel\v{c}i\'{c}, M., Antulov-Fantulin, N., Bo\v{s}njak, M., \v{S}muc, T.:
  Extending rapidminer with recommender systems algorithms.
\newblock In: Proceedings of the RapidMiner Community Meeting and Conference,
  pp. 63--75 (2012)

\bibitem{NetflixAttack}
Narayanan, A., Shmatikov, V.: Robust de-anonymization of large sparse datasets.
\newblock In: Proceedings of the 2008 IEEE Symposium on Security and Privacy,
  SP '08, pp. 111--125. IEEE Computer Society, Washington, DC, USA (2008)

\bibitem{NetworksBook}
Newman, M.: Networks: An Introduction.
\newblock Oxford University Press, Inc. (2010)

\bibitem{Synthetic1}
Raghunathan, T., Reiter, J., Rubin, D.: Multiple imputation for statistical
  disclosure limitation.
\newblock Journal of Official Statistics \textbf{19}(1), 1--16 (2003)

\bibitem{Synthetic4}
Reiter, J.: Inference for partially synthetic, public use microdata sets.
\newblock Survey Methodology \textbf{29}(2), 181--188 (2003)

\bibitem{BPRMF}
Rendle, S., Freudenthaler, C., Gantner, Z., Schmidt-Thieme, L.: Bpr: Bayesian
  personalized ranking from implicit feedback.
\newblock In: Proceedings of the 25th Conference on Uncertainty in Artificial
  Intelligence, UAI '09, pp. 452--461 (2009)

\bibitem{StateArtRS}
Rendle, S., Tso-Sutter, K., Huijsen, W., Freudenthaler, C., Gantner, Z.,
  Wartena, C., Brussee, R., Wibbels, M.: Report on state of the art recommender
  algorithms (update).
\newblock Tech. rep. (2011).
\newblock MyMedia public deliverable D4.1.2.

\bibitem{colFilt2}
Resnick, P., Iacovou, N., Suchak, M., Bergstrom, P., Riedl, J.: Grouplens: an
  open architecture for collaborative filtering of netnews.
\newblock In: Proceedings of the 1994 ACM conference on Computer supported
  cooperative work, CSCW '94, pp. 175--186 (1994)

\bibitem{kAnonymity1}
Samarati, P.: Protecting respondents' identities in microdata release.
\newblock IEEE Transactions on Knowledge and Data Engineering \textbf{13}(6),
  1010--1027 (2001)

\bibitem{kAnonymity2}
Samarati, P., Sweeney, L.: Generalizing data to provide anonymity when
  disclosing information (abstract).
\newblock In: Proceedings of the seventeenth ACM SIGACT-SIGMOD-SIGART symposium
  on Principles of database systems, PODS '98, pp. 188 (1998)

\bibitem{PRAM2}
Wolf, P.P.D., Amsterdam, H.V., Design, C., Order, W.T.: An empirical evaluation
  of pram statistics netherlands voorburg/heerlen (2004)

\bibitem{PRAM1}
Wolf, P.P.D., Gouweleeuw, J.M., Kooiman, P., Willenborg, L.: Reflections on
  pram.
\newblock Statistical Data Protection, Luxembourg pp. 337--349 (1999)

\end{thebibliography}
\end{document}